\documentclass[journal]{IEEEtran}
\UseRawInputEncoding
\pdfoutput=1
\usepackage{amsmath,amsfonts}
\usepackage{algorithmic}
\usepackage{algorithm}
\usepackage{array}
\usepackage[caption=false,font=normalsize,labelfont=sf,textfont=sf]{subfig}
\usepackage{textcomp}
\usepackage{stfloats}
\usepackage{url}
\usepackage{verbatim}
\usepackage{graphicx}
\usepackage{cite}
\usepackage{booktabs}
\usepackage{tabularx}
\usepackage{diagbox} % 加载 diagbox 宏包
\usepackage{bm}
\usepackage{graphics} 
\usepackage{epstopdf} 
\usepackage{multicol}
\usepackage{fancyhdr}

\begin{document}

\title{Channel-robust Automatic Modulation Classification Using Spectral Quotient Cumulants}

\author{Sai~Huang,~\IEEEmembership{ Senior Member,~IEEE,}
	Yuting~Chen,
	Jiashuo~He,~\IEEEmembership{Graduate Student Member,~IEEE,}
	Shuo~Chang,~\IEEEmembership{Member,~IEEE,}
	and~Zhiyong~Feng,~\IEEEmembership{Senior Member,~IEEE}
	\thanks{This work was supported in part by the National Natural Science Foundation
		of China under Grant (62171045, 62201090), in part by the National Key Research and Development Program of China under Grants (2019YFB1804404,
		2020YFB1807602), in part by the BUPT Excellent Ph.D. Students Foundation
		(CX2023233).THIS WORK HAS BEEN SUBMITTED TO THE IEEE FOR POSSIBLE PUBLICATION. COPYRIGHT MAY BE TRANSFERRED WITHOUT NOTICE, AFTER WHICH THIS VERSION MAY NO LONGER BE ACCESSIBLE}
	\thanks{Sai Huang, Yuting Chen, Jiashuo He,  Shuo Chang, and Zhiyong Feng are with the Key Laboratory of
		Universal Wireless Communications, Ministry of Education, Beijing University
		of Posts and Telecommunications, Beijing 100876, China.
		E-mail: \{huangsai, cyting, jiashuohe, changshuo, fengzy\}@bupt.edu.cn.
		(Corresponding author: Jiashuo He.)}}% <-this % stops a space

\maketitle
\thispagestyle{fancy}
\fancyhead[C]{\fontsize{5}{5}\selectfont THIS WORK HAS BEEN SUBMITTED TO THE IEEE FOR POSSIBLE PUBLICATION. COPYRIGHT MAY BE TRANSFERRED WITHOUT NOTICE, AFTER WHICH THIS VERSION MAY NO LONGER BE ACCESSIBLE}

\begin{abstract}
Automatic modulation classification (AMC) is to identify the modulation format of the received signal corrupted by the channel effects and noise. Most existing works focus on the impact of noise while relatively little attention has been paid to the impact of channel effects. However, the instability posed by multipath fading channels leads to significant performance degradation. To mitigate the adverse effects of the multipath channel, we propose a channel-robust modulation classification framework named spectral quotient cumulant classification (SQCC) for orthogonal frequency division multiplexing (OFDM) systems. Specifically, we first transform the received signal to the spectral quotient (SQ) sequence by spectral circular shift division operations. Secondly, an outlier detector is proposed to filter the outliers in the SQ sequence. At last, we extract spectral quotient cumulants (SQCs) from the filtered SQ sequence as the inputs to train the artificial neural network (ANN) classifier and use the trained ANN to make the final decisions. Simulation results show that our proposed SQCC method exhibits classification robustness and superiority under various unknown Rician multipath fading channels compared with other existing methods. Specifically, the SQCC method achieves nearly 90\% classification accuracy at the signal to noise ratio (SNR) of 4dB when testing under  multiple channels but training under  AWGN channel.
\end{abstract}


\section{Introduction}
\IEEEPARstart{R}{ecently}, the Internet of Vehicles (IoV) has developed rapidly with the advent of fifth-generation (5G) mobile unication technology. IoV is a complex network where vehicles are controlled and directed based on the information collected from the sensors installed in them in a real-time manner \cite{sedjelmaci2023secure}. Without effective precautions, intruders or hackers can interfere with the system and make them lose control. This can lead to economic losses in terms of damage to vehicles and even endanger the lives of drivers and passengers. Hence, it is necessary to detect the attacks and ensure the physical layer security. Automatic modulation classification (AMC) is a promising way to struggle against the potential IoV physical layer threats as it can detect and identify the modulation type of the attack signals \cite{wang2019data,hanna2021signal}. Orthogonal frequency division multiplexing (OFDM)  is a widely known multicarrier modulation technology utilized in the high spectral efficiency wireless communication, and hence the modulation classification of OFDM signals is a significant research challenge for ensuring normal communication progress. 

Generally speaking, AMC methods can be divided into two categories: likelihood-based (LB) methods and feature-based (FB) methods, respectively. The LB method considers modulation classification as a multiple hypotheses testing problem using the derived likelihood function of the received signal \cite{zhang2017cooperative}. LB can perform the optimal performance in a Bayesian sense, but it requires intensive computational complexity and  some prior knowledge of the signal parameters \cite{9042348}. In contrast to the LB method, the FB method extracts the statistical features of the received signal and generally performs a sub-optimal performance with low computational complexity, and hence it is widely utilized in practice \cite{s22031020}. The most extracted features include cyclostationarity \cite{dobre2011cyclostationarity}, high-order cumulants \cite{wang2010fast,837045,ALI2020173,6060143,8911517}, and Fourier transforms \cite{4678248}. For instance, Swami \textit{et al.} in \cite{837045} first employed the fourth-order cumulants to discriminate the modulation types through the threshold-based classifier, and it is robust to phase mismatches. However, the performance largely degrades when to classify higher-order M-ary quadrature amplitude modulation (M-QAM) types. Hence, an improved method was proposed in \cite{ALI2020173}, which employs a novel classifier based on logarithmic functions to classify features. However, this algorithm is only applicable to additive white Gaussian noise (AWGN) channel, and its performance will fall below 40\% under the multipath fading channels since the statistics-based features are highly sensitive to the distribution shifts resulting from the variant channel environments. 

To pursue the channel-robust performance, Shih \textit{et al.} in  \cite{6060143} proposed a blind channel coefficient estimation algorithm combined with high-order statistics (HOS) to combat the effect of multipath fading, and this method performs well even in four-tap multipath. However, this approach is designed with the criterion that the power of the line-of-sight (LOS) path is much larger than the total power of the rest paths. Xing \textit{et al.} in \cite{8911517} proposed a logarithmic functional fitting (LFF)-based AMC method in the single-carrier system, where the cepstrum is used to release the multipath fading effects. However, this approach requires multi-step complex operations in the signal preprocessing module, which makes it difficult to apply in practice. Moreover, the performance degrades when considering the classification of the OFDM wideband signals. Gupta \textit{et al.} in \cite{9042348} extracted fourth-order cumulants from the discrete Fourier transform (DFT) of the received orthogonal frequency division multiplexing (OFDM) signal along with its squared signal. However, this approach cannot be applied in practice due to its lack of ability to AMC when the channel state changes. In summary, there is still a critical need  to perform robust classification with unknown multipath fading channels.

In a recent study \cite{he2023radio}, it is established that the spectral quotient (SQ) sequence, acquired through spectral cyclic shift division (SCSD) of an OFDM symbol, demonstrates remarkable robustness to channel effects. The authors utilize the proposed SQ sequence of OFDM signal to identify the transmitters under time-varying multipath channels. However, directly using the SQ sequence in the AMC task leads to a deterioration in classification performance due to the outliers presented in this sequence. These findings inspire us to enhance AMC performance in multipath channels by harnessing features extracted from the transformed SQ sequence.
%we find that the classification performance of AMC  degrades when directly using the SQ sequence. 

In this paper, a channel-robustness OFDM modulation classification framework using spectral quotient cumulants (SQCs) feature is proposed. Specifically, we first propose an outlier removal detector to filter the outliers in the SQ sequence which is transformed from the received signal by SCSD operations. Secondly, we extract the SQCs  from the  filtered SQ sequence. Finally, an artificial neural network (ANN) classifier trained under AWGN channel is utilized to identify the modulation types of the received OFDM signals under multipath fading channels. Experimental results demonstrate the superior performance and the robustness of our proposed spectral quotient cumulant classification (SQCC) method under different unknown multipath fading channels.  
\begin{figure*}[t]
	\includegraphics[scale=0.51]{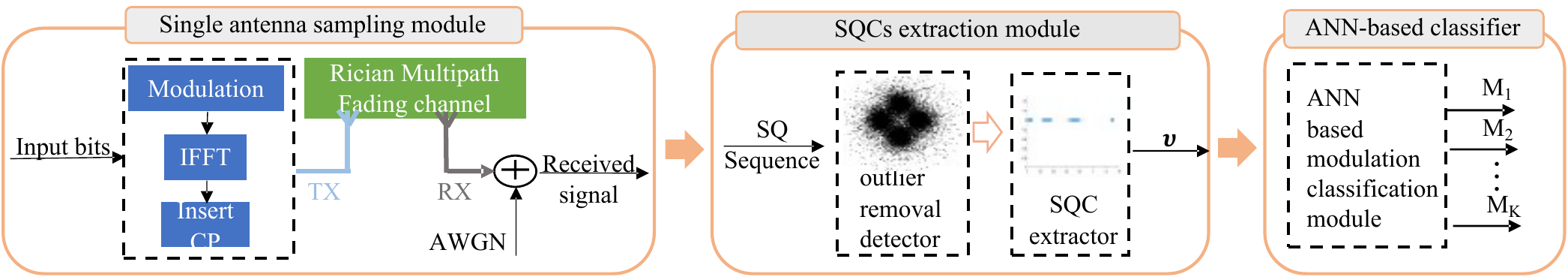}
	\caption{The proposed SQCC framework.}
	\label{overview}        		% 引用标签
	\vspace{-0.3cm} %设置与上面正文的距离
\end{figure*}
\section{System Framework and Signal Model}
\label{section:signalmodel} %这里是章节的标签，引用时需要
\subsection{System Framework}
As shown in Fig. \ref{overview}, the SQCC framework can be roughly divided into three modules, i.e., single antenna sampling module, SQCs feature extraction module, and ANN-based classifier module. The single antenna sampling module first receives the OFDM signal and transforms the received signal transmitted by the jammer into the baseband one. In the SQCs feature extraction module, we propose an outlier removal detector to filter the outliers in the SQ sequence and then extract SQCs from the filtered SQ sequence.  In the last ANN-based classifier module, the well-trained ANN-based classifier obtains $K$ score values associated with each modulation type, and the final predictions of the modulation types are made by adopting the hypothesis with the highest score.

\subsection{Signal Model}
Considering a baseband OFDM system detailed in the single antenna sampling module in Fig. \ref{overview}, the received signal through  Rician multipath fading channel is given as\\
\begin{equation}
	\begin{aligned}
		y^{k}(n) = \sum_{l = 0}^{L-1} h_{l} x^{k}(n-\tau_{l})+w(n) , \\
		\quad 0 \leq n \leq N-1 ,
	\end{aligned}	
\end{equation}
where $N$ is the received signal length; $k$ indicates the $k_{th}$ modulation type; $w(n)$ is the AWGN with zero mean and variance of  $\sigma^{2}$. $h_{l}$ denotes the $l_{th}$ channel coefficient and $L$ is the total number of channel delay taps. The Rician channel model assumes the existence of a dominant LOS path in addition to the scattered path [13]. In general, the level of fading severity in the first tap can be qualified by Rician $K$-factor, which is defined as $K_{f}=P_{LOS}/P_{r} $, where $P_{LOS}$ denotes the power of LOS component and $P_{r}$ denotes the power of the complex Gaussian variable.  $x^{k} (n)$ stands for the transmitted OFDM signal after performing  the inverse fast Fourier transform (IFFT), which can be written as\\
\begin{equation}
	\begin{aligned}
		x^{k}(n) = \frac{1}{N}\sum_{m = 0}^{N-1}X^{k}(m)e^{j2\pi mn/N};\\ 
		\quad -N_{cp} \leq n \leq N-1,
	\end{aligned}	
\end{equation}
where $X^{k}(m)$ is the modulated data using modulation type of $M_{k}$. The cyclic prefix (CP) of length $N_{cp}$ is appended ahead of each symbol to mitigate the intersymbol interference (ISI). 
%For the received signal $y^{k}(n)$, the task of AMC is to distinguish its modulation type. 

\section{Spectral Quotient Cumulant Classification} 
In this section,  the proposed outlier removal detector to filter outliers is firstly illustrated.  Secondly, the definition of the proposed SQCs is given and the theoretical values of different modulation format are illustrated. Finally, ANN is trained using the filtered SQ sequence under AWGN channel and is utilized to make the final decisions under the multipath fading channels.
\subsection{Signal Preprocessing}
To mitigate the channel effect, the data transformation of the received signal $y^{k}(n)$ to the SQ sequence $\bm{\Upsilon}$ is adopted according to \cite{he2023radio}, which is given by
\begin{equation}
	\begin{aligned}
		\Upsilon(m)=\frac{Y^{k}(m)}{Y^{k}(m_{d})}, 
		\quad m_{d}=\left\{\begin{array}{cc}
			N-1+m, & m<1 \\
			m-1, & m \geq 1
		\end{array}\right. \,,
	\end{aligned}	
\end{equation}
where $\Upsilon(m)$ denotes the $mth$ element of the generated SQ sequence $\mathbf{\Upsilon}=[\Upsilon(0),\dots,\Upsilon(m),\dots,\Upsilon(N-1)]^{T}$ by carrying out the SCSD algorithm with one-step shift. $Y^{k}(m)$ is the received OFDM symbol by performing the fast Fourier transform (FFT) on the received signal $y^{k}$ as  
\begin{equation}
	\begin{aligned}
		Y^{k}(m) = \sum_{n = 0}^{N-1}y^{k}(n)e^{-j2\pi nm/N},\\ 
		\quad 0 \leq m \leq N-1 .
	\end{aligned}	
\end{equation}

Since the division operation of spectral circular shift division (SCSD) is sensitive to the denominator values, the outliers are likely to be generated in the SQ sequence. This can significantly destabilize the statistical properties and thus degrade the classification performance. Therefore, we propose an outlier removal detector to filter the outliers in the SQ sequence $\bm{\Upsilon}$, where each point is classified by its values and then the filtered SQ sequence $\mathbf{\Gamma}$ can be constructed as 
\begin{align}\label{filter}
	\mathbf{\Gamma}=\left \{\Gamma\in \bm{\Upsilon} | \Gamma<= T_{max} \right \}, 
\end{align}	
where $T_{max}$ denotes the specified maximum threshold. 
The procedure can be described in the following steps.

\begin{itemize}
	\item \textit{Step 1 :} Determining the Maximum Threshold ($T_{max}$) 
	
	We present a rigorous method for determining $T_{max}$ based on validation results. This threshold value will serve as a crucial reference point for outlier identification.
	
	\item \textit{Step 2 :} Identifying and Removing Outliers
	
	For each element in $\mathbf{\Upsilon}$, a systematic comparison is conducted with $T_{max}$. Any elements exceeding $T_{max}$ are identified as outliers and subsequently removed from $\mathbf{\Upsilon}$.
	%the SQ sequence
	
	\item \textit{Step 3 :} Constructing the Filtered SQ Sequence
	
	Following the removal of outliers in Step 2, the filtered SQ sequence  $\mathbf{\Gamma}$  is constructed using the remaining elements. This refined sequence is expected to exhibit enhanced statistical properties, facilitating more accurate classification.
\end{itemize} 

\subsection{Definitions and Sample Estimates of SQCs}
\label{section:SQC}
\begin{table}[!t]
	\caption{Theoretical Cumulants Based Spectral Quotient \label{cum}}
	\centering
%	\vspace{-5pt} %设置与上面正文的距离
	\begin{tabular}{c|ccccc}
		\hline
		SQC & BPSK & QPSK & 8PSK & 8QAM & 16QAM   \\
		\hline
	\textit{$C_{20}$} & 1 & 0 & 0 & 0.16 & 0 \\
	\textit{$C_{21}$} & 1 & 1 & 1 & 1.8 & 1.89 \\
	\textit{$C_{22}$} & 0 & 0 & 0 & 0.13 & 0  \\
	\textit{$C_{40}$} & 2 & 1 & 0 & 1.56 & 4.20  \\
	\textit{$C_{41}$} & 2 & 0 & 0 & 0.704 & 0 \\
	\textit{$C_{42}$} & 2 & 1 & 1 & 0.25 & 1.876  \\
		\hline
	\end{tabular}
\end{table}
For a SQ signal $\Gamma(m)$, the general definition of SQCs can be given by
\begin{flalign}
	C_{\alpha,\beta}(\Gamma)&=\sum\left [(-1)^{(P-1)}(P-1)! \prod_{p=1}^{P}M_{\alpha_{p},\beta_{p}}(\Gamma)   \right ],
\end{flalign}	
where $M_{\alpha_{p},\beta_{p}}(\Gamma)=E \left \{ \Gamma(m)^{\alpha_{p}-\beta_{p}}\cdot(\Gamma^{*}(m))^{\beta_{p}} \right \}$ is the  $\alpha_{p}$ order mixing moment of $\Gamma(m)$ with $\beta_{p}$ conjugate and $E\left \{ \cdot \right \} $ outputs the expectation. The external summation iterates through all partitions $\left \{ 1,\dots,\alpha \right \}$  and each partition has $P$ moments of $\alpha_{p}$ order and  $\beta_{p}$ conjugate. 
Here, we use the SQ sequence $\mathbf{\Gamma}$ containing $N$ elements to estimate the moment, which is given by
\begin{align}
	&\hat{M}_{\alpha ,\beta }(\Gamma)=\frac{1}{N}\sum_{m=1}^{N}\left ( \Gamma(m)^{\alpha -\beta } \cdot (\Gamma^{\ast}(m)^{\beta} \right ), &
\end{align}	
where the superscript $\wedge$ refers to a sample average estimate. Then we obtain the following sample estimates of SQCs as 
\begin{equation}
\begin{aligned}\label{1}
\hat{C}_{20}&=\frac{1}{N}\sum_{m=1}^{N} \Gamma^{2}(m), \\ 
\hat{C}_{21}&=\frac{1}{N}\sum_{m=1}^{N}|\Gamma(m)|^{2},\\ 
\hat{C}_{22}&=\frac{1}{N}\sum_{m=1}^{N}\Gamma^{\ast}(m)^{2}-\hat{C}_{20}^{2}-2*\frac{1}{N}\sum_{m=1}^{N}\Gamma(m),\\
\hat{C}_{40}&=\frac{1}{N}\sum_{m=1}^{N}\Gamma^{4}(m)-3\hat{C}_{20}^{2},\\
\hat{C}_{41}&=\frac{1}{N}\sum_{m=1}^{N}\Gamma(m)^{3} \Gamma^{\ast}(m)-3\hat{C}_{20}\hat{C}_{21},\\
\hat{C}_{42}&=\frac{1}{N}\sum_{m=1}^{N}|\Gamma(m)|^{4}-|\hat{C}_{20}|^{2}-2\hat{C}_{21}^{2}.
\end{aligned}
\end{equation}
%\begin{align}\label{1}
%	&\hat{C}_{20}=\frac{1}{N}\sum_{m=1}^{N} \Gamma^{2}(m), &
%\end{align}	
%\begin{align}\label{2}
%	&\hat{C}_{21}=\frac{1}{N}\sum_{m=1}^{N}|\Gamma(m)|^{2}, &
%\end{align}	
%\begin{align}\label{3}
%	&\hat{C}_{22}=\frac{1}{N}\sum_{m=1}^{N}\Gamma^{\ast}(m)^{2}-\hat{C}_{20}^{2}-2*\frac{1}{N}\sum_{m=1}^{N}\Gamma(m), &
%\end{align}	
%\begin{align}\label{4}
%	&\hat{C}_{40}=\frac{1}{N}\sum_{m=1}^{N}\Gamma^{4}(m)-3\hat{C}_{20}^{2}, &
%\end{align}	
%\begin{align}\label{5}
%	&\hat{C}_{41}=\frac{1}{N}\sum_{m=1}^{N}\Gamma(m)^{3} \Gamma^{\ast}(m)-3\hat{C}_{20}\hat{C}_{21},&
%\end{align}	
%\begin{align}\label{6}
%	&\hat{C}_{42}=\frac{1}{N}\sum_{m=1}^{N}|\Gamma(m)|^{4}-|\hat{C}_{20}|^{2}-2\hat{C}_{21}^{2}. &
%\end{align}	

Tab. \ref{cum} lists the theoretical values of SQCs for candidate modulation types.
It should be noted that these theoretical values are derived from Eq. (\ref{1}) under a noise-free scenario, where the transmitted OFDM symbols are equiprobable.
%Eq.(\ref{1})-(\ref{6})
\subsection{ANN-based Classifier}
\label{subsection:classifier} %这里是章节的标签，引用时需要
With the extracted SQCs from the SQ sequence, a SQC vector $\boldsymbol{\nu} = [\hat{C}_{20}, \hat{C}_{21}, \hat{C}_{22}, \hat{C}_{40}, \hat{C}_{41}, \hat{C}_{42}]^{T}$ standing for the unknown modulation type is constructed, where the values of $\boldsymbol{\nu}_{i}$ is related to specific signal-to noise ratio (SNR). Then we utilize a simple ANN, consisting of an input layer, hidden layers, and output layer to complete the classification task. The feature vector  $\boldsymbol{\nu}$ serves as the input of the network. The last layer employs the softmax activation function to normalize the output values of the output layer, where the output of each node indicates the probability that the feature vector  $\boldsymbol{\nu}$ is related to the corresponding modulation type. Moreover, we apply the early stopping method \cite{prechelt2002early} on the validation set to avoid overfitting.

Cross-entropy loss is employed as the loss function to train the network. The cross-entropy loss is given as
\begin{equation}
	\begin{aligned}
		L = -\frac{1}{N_{s}}\sum_{i=1}^{N_{s}}\sum_{k=1}^{K}t_{ik}log(P(\hat{M}_{mod} = M_{k}|y^{i})) ,
	\end{aligned}	
\end{equation}
where $N_{s}$ and $K$ are the number of training samples and the set of candidate modulation types, respectively. $t_{ik}$ represents the ground truth label of the $i_{th}$ sample corresponded to the $k_{th}$ modulation type and $P(\hat{M}_{mod} = M_{k}|y^{i})$ represents the probability that the sample $y_{i}$ is predicted by the network to be class $H_{k}$. The training objective is to minimize the loss $L$ by optimizing the network parameters. When we get the optimal parameters of ANN, the modulation classification is given by referring to the maximum-a-posterior (MAP) criterion
\begin{equation}
	\begin{aligned}
		\hat{M_{k}} = \arg\max_{1\le k \le K}P(\hat{M}_{mod} = M_{k}|y^{M_{k}}) ,
	\end{aligned}	
\end{equation}
where $\hat{M_{k}}$ is the predicted modulation type and $P(\hat{M}_{mod} = M_{k}|y^{M_{k}})$ is the probability when the modulation type is correctly classified as $M_{k}$.

\section{Simulations}

\begin{table}[t]
	\caption{The Power Delay Profiles of Rician Multipath Channels}
	\label{pdp}       % Give a unique label
	\centering
	\vspace{-5pt} %设置与上面正文的距离
	\begin{tabular}{c|ccccc}
		\toprule
		\diagbox{Path Delay}{Power}{Channel Code} & \textit{$\mathcal{H}_{1}$} & \textit{$\mathcal{H}_{2}$}  \\
		\hline
		$\tau _{0}$  & 0.75  & 0.55 \\
		$\tau_{1}$ & 0.25 & 0.45 \\
		\bottomrule
	\end{tabular}%
\end{table}
In this section, we will evaluate the performance of the proposed AMC algorithm, in terms of the average probability of correct classification ($P_{cc}$). Assuming there are $K$ candidate modulation types, then $P_{cc}$ can be formulated as
\begin{equation}
	\begin{aligned}
		P_{cc} = \sum_{k = 1}^{K} P(\hat{M}_{mod} = M_{k}|y^{M_{k}})P(y^{M_{k}}),
	\end{aligned}	
\end{equation}
where $P(y^{M_{k}})$ is the prior probability of $M_{k}$.

For the following evaluation, we consider two scenarios of different modulation type candidate pools, namely $\mathbf{\Theta_{1}} = \left \{ \text{BPSK}, \text{QPSK}, \text{16QAM} \right \} $  and  $\mathbf{\Theta_{2}} = \left \{ \text{BPSK}, \text{QPSK}, \text{8PSK}, \text{8QAM},\text{16QAM} \right \} $. $\mathbf{\Theta_{1}}$ is applicable for most algorithm, whereas $\mathbf{\Theta_{2}}$ is more challenging. The length of OFDM signal and CP are set to 256 and 64. AWGN channel is denoted as $\mathcal{H}_{0}$. Two Rician multipath fading channels (i.e., $\mathcal{H}_{1}$, $\mathcal{H}_{2}$) are considered, where the $K_{f}$ is set to 9 and the detailed power delay profiles are given in Tab. \ref{pdp}.  Moreover, the channel coefficients are randomly and periodically generated for each signal frame. For each modulation type, we generate 800 signal frames for each SNR in a range [-10, 10] dB (step is 2 dB) under different channel conditions, where each signal frame contains 80 OFDM symbols. 

\subsection{Ablation Study}
\begin{figure}[t]
	%是可选项 h表示的是here在这里插入，t表示的是在页面的顶部插入
	\centering
	\vspace{-0.3cm} %设置与上面正文的距离
	\includegraphics[scale=0.24]{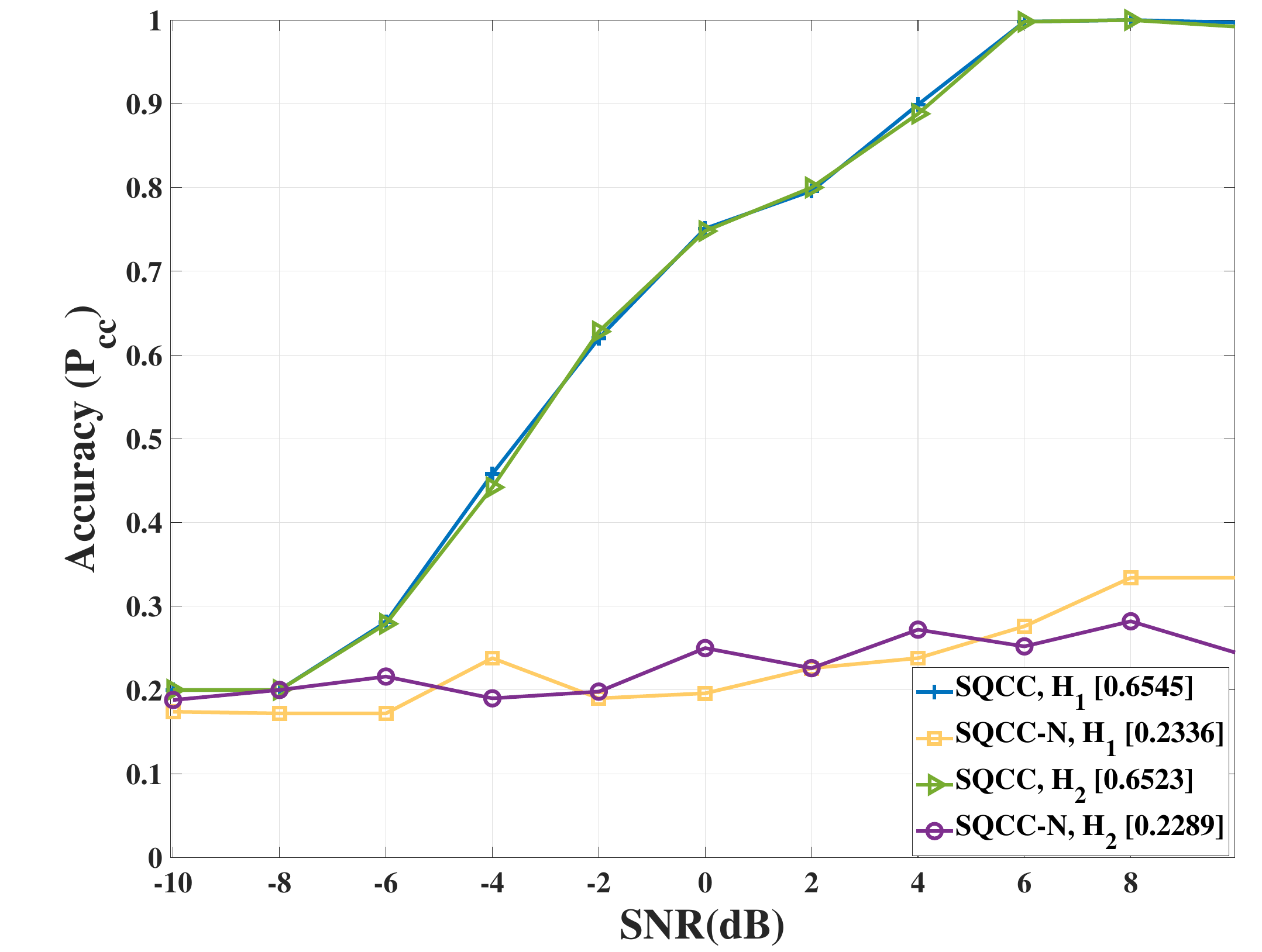}
	\caption{The ablation study results under different multipath fading channels }
	\label{ablation}
\end{figure}
To evaluate the effectiveness of the proposed SQ signal preprocessing scheme on the overall classification accuracy, Fig. \ref{ablation} illustrates the classification results with and without the outlier removal detector, which are marked as SQCC and SQCC-N, respectively.
It is clear that the proposed SQCC method outperforms the SQCC$\mbox{-}$N method along all effective SNRs under both $\mathcal{H}_{1}$ and $\mathcal{H}_{2}$. Obviously, SQCC$\mbox{-}$N even utterly fails as the result of a random selection among candidates, which infers that the outliers can severely pollute the statistical properties. 
With the aid of the outlier removal detector, the overall classification accuracy can be improved at least 41.87$\%$.
Hence, we can draw a conclusion that the outlier removal detector is effective to improve the classification performance.

\subsection{Performance Comparison}
\begin{figure}[t]
	\centering
	\vspace{-0.3cm} %设置与上面正文的距离
	\includegraphics[scale=0.24]{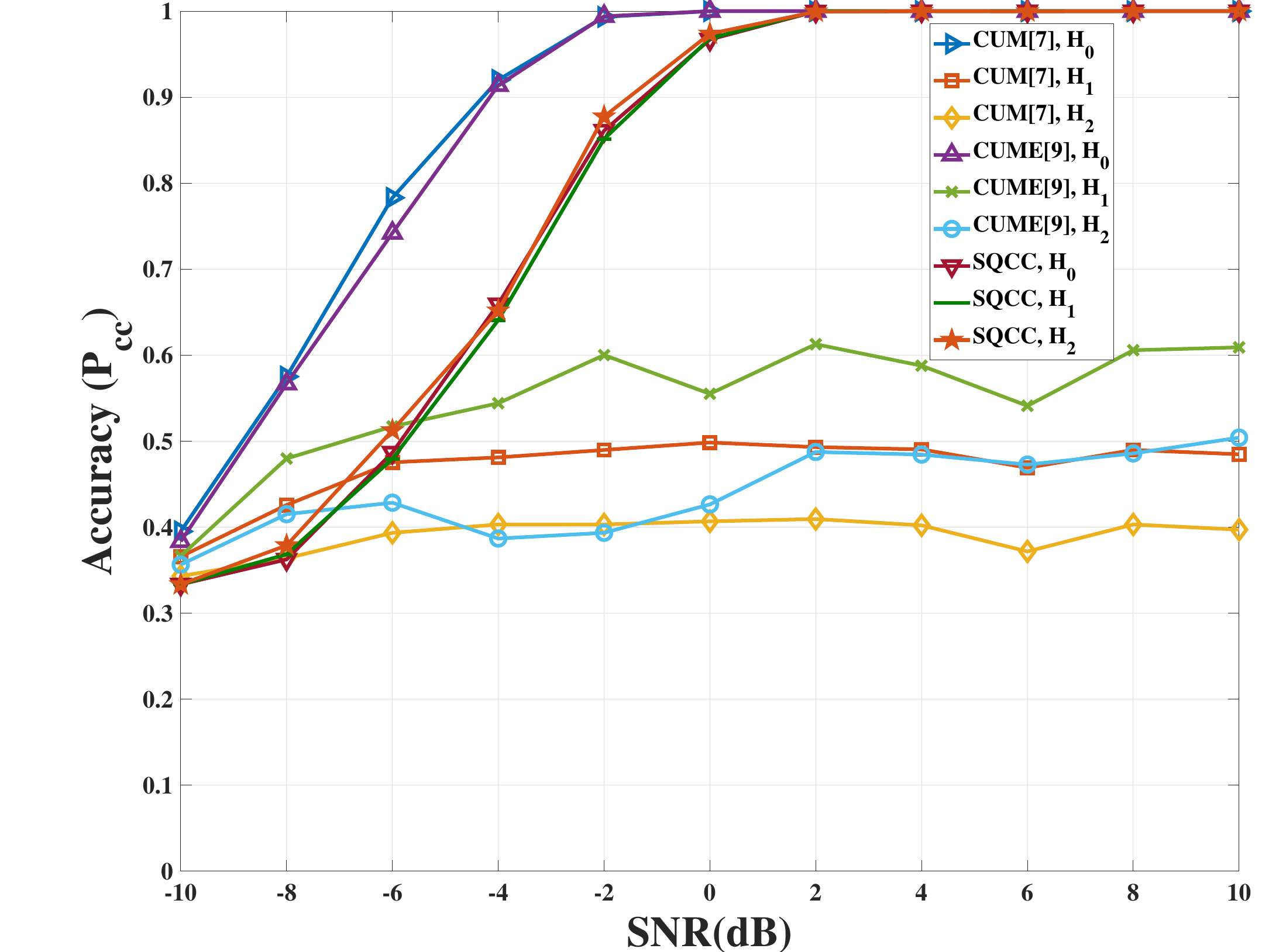}
	\caption{The comparison results of different methods in $\mathbf{\Theta_{1}}$ under different channels at different SNRs.}
	\label{exp1_compare}        		% 引用标签
\end{figure}
\begin{figure}[t]
	\centering
	\vspace{-0.3cm} %设置与上面正文的距离
	\includegraphics[scale=0.24]{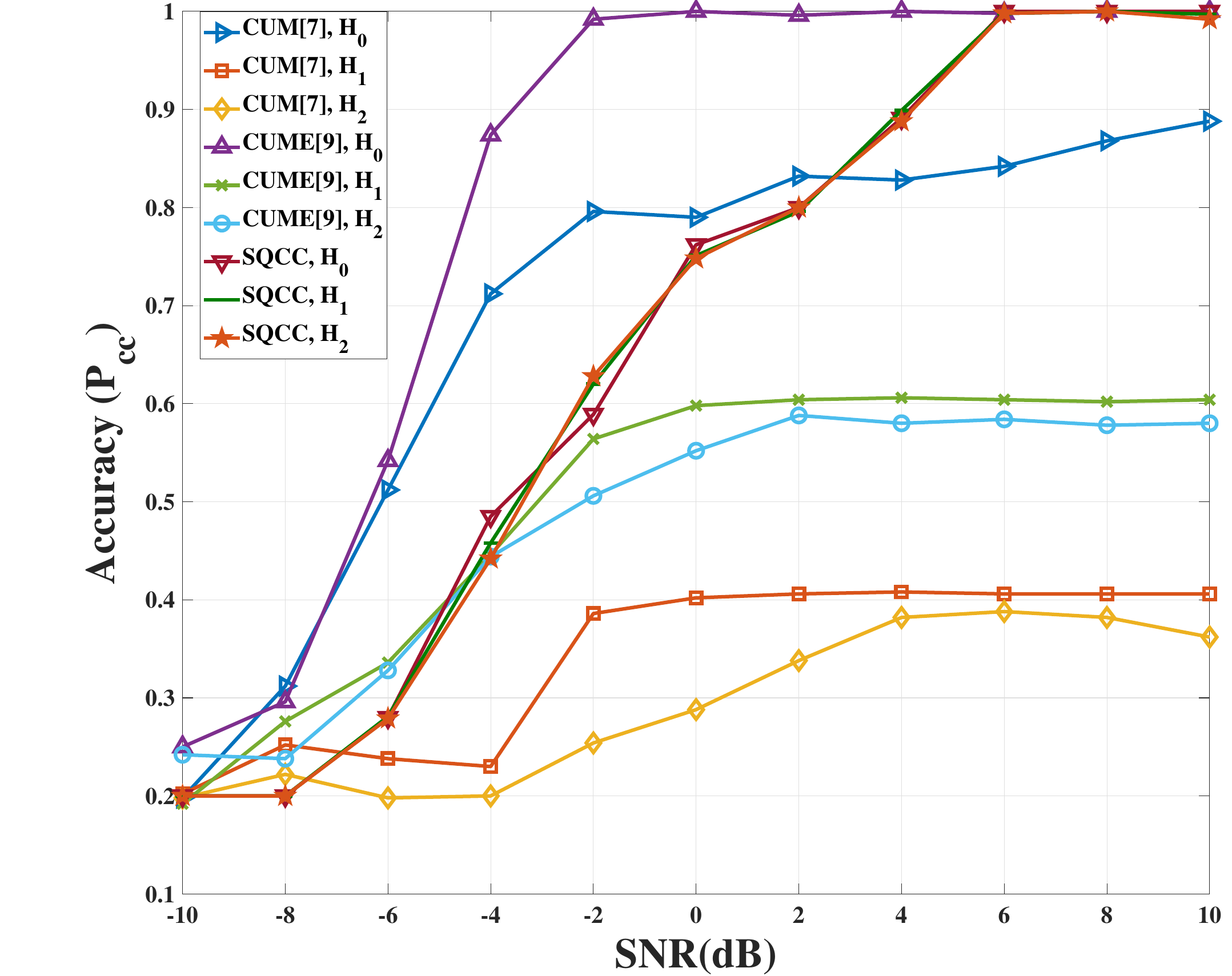}
	\caption{The comparison results of different methods in $\mathbf{\Theta_{2}}$ under different channels at different SNRs.}
	\label{exp2_compare}        		% 引用标签
\end{figure}
In this subsection, we compare the classification accuracy of our method with two other existing FB-AMC methods, i.e., CUM\cite{837045} and CUME\cite{6060143}. In general, ANN classifiers perform better than decision tree classifiers and threshold classifiers. Hence, to avoid the effect of different classifiers on classification accuracy, we replace the classifiers of the competitors with the ANN classifier designed in Section \ref{subsection:classifier} in the following comparison simulations.   

Fig. \ref{exp1_compare} shows the classification performance in terms of the classification accuracy on $\mathbf{\Theta_{1}}$ in three different channels (i.e., $\mathcal{H}_{0}$, $\mathcal{H}_{1}$, $\mathcal{H}_{2}$). 
It can be observed that the competitors outperform our method at low SNR levels over AWGN channel. 
When the SNR is larger than 2 dB, our method can achieve the same peak accuracy of 100$\%$ as these competitors.
However, the classification performance of both CUM\cite{837045} and CUME\cite{6060143} degrades significantly when the channel conditions are changed during the testing stage.
Specifically, for Rician channel cases, the peak accuracies of CUM\cite{837045} and CUME\cite{6060143} drop at least 40$\%$ and 50$\%$ when compared to the results under AWGN channel.
Moreover, from the curves marked with $\mathcal{H}_{1}$ and $\mathcal{H}_{2}$, we can find the drop is more significant when the fading effect becomes more severe.
As for the proposed SQCC method, it remains the similar performance under different channel conditions, which means that our method is robust to the channel variations.

In addition, Fig. \ref{exp2_compare} shows  the classification results in $\mathbf{\Theta_{2}}$. 
It can be appreciated that the CUM\cite{837045} only reaches about 90$\%$ peak accuracy at SNR = 10 dB under AWGN channel.
This can be attributed to that the CUM\cite{837045} method can not classify the \text{QPSK} and \text{8QAM} well.
The CUME\cite{6060143} also achieves an excellent classification performance over AWGN channel, reaching 100$\%$ accuracy at SNR = -2 dB.
Again, for Rician multipath channels, these competitors performs badly, with a significant drop of at least 40$\%$ in peak accuracy.
As for our method, it can classify these modulation types with an accuracy of 100$\%$ at 6 dB in each case.
Thus, we can conclude that our method is effective to classify the modulation types in $\mathbf{\Theta_{1}}$ and $\mathbf{\Theta_{2}}$ when encountered the unknown channel conditions.

\section{Conclusion}
In this correspondence, we proposed a novel SQCC framework to overcome the drawback of existing FB-AMC methods, where the features extracted directly from the received signal would vary in multipath fading channels. The proposed SQCs were constructed from the transformed SQ sequence, which contained discriminant and channel-robust information for different modulation types. The simulation results clearly demonstrate the robustness and superiority of the proposed AMC method in comparison to the recent methods when the unknown channel conditions are considered.

%\section*{Acknowledgments}
%This should be a simple paragraph before the References to thank those individuals and institutions who have supported your work on this article.
%
%
%
%{\appendix[Proof of the Zonklar Equations]
%Use $\backslash${\tt{appendix}} if you have a single appendix:
%Do not use $\backslash${\tt{section}} anymore after $\backslash${\tt{appendix}}, only $\backslash${\tt{section*}}.
%If you have multiple appendixes use $\backslash${\tt{appendices}} then use $\backslash${\tt{section}} to start each appendix.
%You must declare a $\backslash${\tt{section}} before using any $\backslash${\tt{subsection}} or using $\backslash${\tt{label}} ($\backslash${\tt{appendices}} by itself
% starts a section numbered zero.)}


 \small

\end{document}